\documentclass[preprint,aps,eqsecnum,graphicx]{revtex}

\def\thetap{\hbox{$\Theta^+$}}
\def\beq#1{\begin{equation} \label{#1}}
\def\eeq{\end{equation}}
\def\URLtilde{\lower0.2em\hbox{$\tilde{\phantom{a}}$}}

\catcode`\@=11 
\def\lsim{\mathrel{\mathpalette\@versim<}}
\def\gsim{\mathrel{\mathpalette\@versim>}}
\def\@versim#1#2{\vcenter{\offinterlineskip
        \ialign{$\m@th#1\hfil##\hfil$\crcr#2\crcr\sim\crcr } }}
\catcode`\@=12 

\def\sbar{\hbox{$\bar s$}}

\newcommand{\bea}{\begin{eqnarray}}
\newcommand{\eea}{\end{eqnarray}}
\def\bra#1{\left\langle #1\right\vert}
\def\ket#1{\left\vert #1\right\rangle}
\def\epsp{\epsilon^{\prime}}
\def\NPB{{ Nucl. Phys.} B}
\def\PLB{{ Phys. Lett.} B}
\def\PRL{ Phys. Rev. Lett.}
\def\PRD{{ Phys. Rev.} D}

\begin{document}
{
\tighten
\preprint {\vbox{
 \hbox{TAUP 2910/10}
 \hbox{WIS/04/10-FEB-DPPA}
}}

\title {About a Possible Nonstrange Cousin of the $\Theta^+$ Pentaquark}
\author{Marek Karliner\,$^{b\,1}$
\\
and\\
Harry J. Lipkin\,$^{a,b,c,\,2}$\,\thanks{Invited talk at International
Workshop {\em ``Narrow Nucleon Resonances:
Predictions, Evidences, Perspectives"}, Edinburgh, Scotland,
June 8--10, 2009.\hfill\break
Supported in part by
U.S.
Department of Energy, Office of Nuclear Physics, under contract
number
DE-AC02-06CH11357.}}
\address{ \vbox{\vskip 0.truecm}
  $^a$Department of Particle Physics
  Weizmann Institute of Science, Rehovot 76100, Israel \\
\vbox{\vskip 0.truecm}
$^b$Raymond and Beverly Sackler School of Physics and Astronomy
\\
Tel Aviv University, Tel Aviv, Israel  \\
\vbox{\vskip 0.truecm}
$^c$Physics Division, Argonne National Laboratory,
Argonne, IL 60439-4815, USA\\
~\\
$^1${\tt marek@proton.tau.ac.il}
\\
$^2${\tt harry.lipkin@weizmann.ac.il}
\\~\\
}

\maketitle

\begin{abstract}
We discuss the implications of the suggested interpretation  of the
recently
reported narrow $\pi N$ resonance (width $\approx 25$ MeV at 1680 MeV)
as a pentaquark in the same multiplet as the $\Theta^+$.
We consider a diquark-triquark pentaquark model involving a recoupling of
the five quarks into a diquark-triquark system in non-standard color
representations. We estimate the mass using
a well-tested simple mass
formula.
Our rough numerical estimate puts the $\pi N$ pentaquark resonance
at 1720 MeV, sufficiently close to the
reported value of 1680 MeV to indicate that this
approach deserves further more accurate investigation.

\end{abstract}

} 


\def\beq#1{\begin{equation} \label{#1}}
\def\eeq{\end{equation}}
\def\bra#1{\left\langle #1\right\vert}
\def\ket#1{\left\vert #1\right\rangle}
\def\epsp{\epsilon^{\prime}}
\def\NPB{{ Nucl. Phys.} B}
\def\PLB{{ Phys. Lett.} B}
\def\PRL{ Phys. Rev. Lett.}
\def\red{
}
\def\black{
}

\def\mycomm#1{\hfill\break\strut\kern-3em{\red\tt ====> #1\black}\hfill\break}
\def\mystrut{\vrule height 6.0ex depth 0.2ex width 0pt}
\def\PRD{{ Phys. Rev.} D}
\section{The reported narrow $\pi N$ resonance}
The reported narrow $\pi N$ resonance (width $\approx$ 25 MeV at 1680 MeV
suggests \cite{Arndt:2003ga}
that it might be a nonstrange pentaquark
in same $SU(3)_f$ multiplet as the antistrange $\Theta^+$ pentaquark with
mass 1540 MeV. The experimental discovery \cite{Nakano:2003qx} of the
\thetap\ with  a
very small \hbox{width~$\lsim 20$ MeV} and a presumed quark configuration $uudd\sbar$
had given rise to a number of further experiments \cite{hicks} and an
interest in theoretical models \cite{jenmalt} for exotic hadrons including
models with diquark structures \cite{NewPenta}.

The controversy between
experimental evidence for and against the existence \cite{cryptopen,ichepproc}
of the \thetap\ remains unresolved. There are also questions about isospin
asymmetry \cite{Hosaka}. The ball now is in the experimental court. New data are needed
to pin down whether these states are really there and to give clues regarding
their structure.

\strut\vskip-2cm\strut
\subsection {Two proton-neutron asymmetries with different $SU(3)_f$ explanations}

Both pentaquark candidates are seen in photoproduction from a
deuteron. Both are reported to be photoproduced much more strongly
on neutrons than on protons. The two theoretical mechanisms proposed for this
asymmetry are very different but both assume
$SU(3)_f$ symmetry. Both are violated by $SU(3)_f$ symmetry breaking.

The discrepancy between  photoproduction of the reported narrow $\pi N$
resonance on protons vs. neutrons  follows directly from an $SU(3)_f$ selection rule if both candidates are
classified in the same $SU(3)$ antidecuplet ($\overline{\bf 10}$).
The analysis is simplified by introducing the ``$U$-spin" SU(2) subgroup
of $SU(3)_f$\cite{Uspin},
analogous to isospin. $U$-spin interchanges $d$ and $s$ quarks, just like
isospin interchanges $u$ and $d$ quarks. In the $SU(3)_f$ limit isospin
and $U$-spin are equally valid. Since $d$ and $s$ quarks both have the same
electric charge (-1/3), all members of the same $U$-spin multiplet have the
same electric charge.

The positively charged members of  the ($\overline{\bf 10}$) multiplet all have $U$-spin 3/2;
the neutral members have $U$-spin 1. The photon has $U$-spin zero
\cite{UspinPhoton},
the proton has
$U$-spin 1/2 and the neutron has $U$-spin 1, as illustrated in Fig.~1.
\eject

\vbox{
\strut
\hfill\break
\hfill\break
\centerline{
\includegraphics[width=20em,clip=true,angle=90]{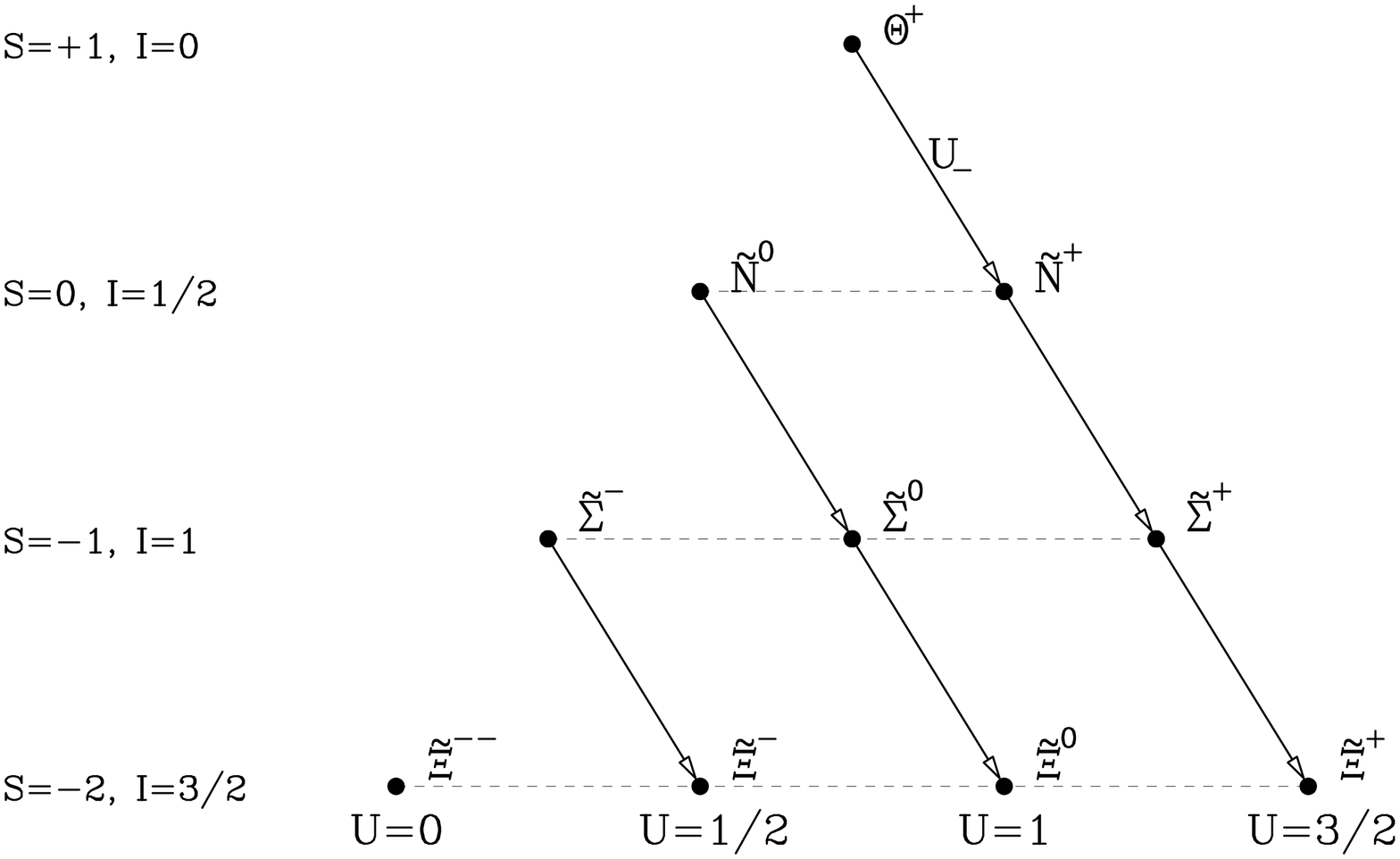}
}
\vskip0.5cm
{\small \em
Fig. 1 (a) Exotic baryon anti-decuplet. The positively charged states
$\Theta^+$, $\tilde N^+$, $\tilde\Sigma^+$ and \ $\tilde\Xi^+$
form a $U$-spin multiplet with $U={3\over2}$. The neutral states
$\tilde N^0$, $\tilde\Sigma^0$ and $\tilde\Xi^0$ form a $U$-spin triplet
with $U=1$.
\hfill\break
}
\vskip1cm
\centerline{
\includegraphics[width=12.5em,clip=true,angle=90]{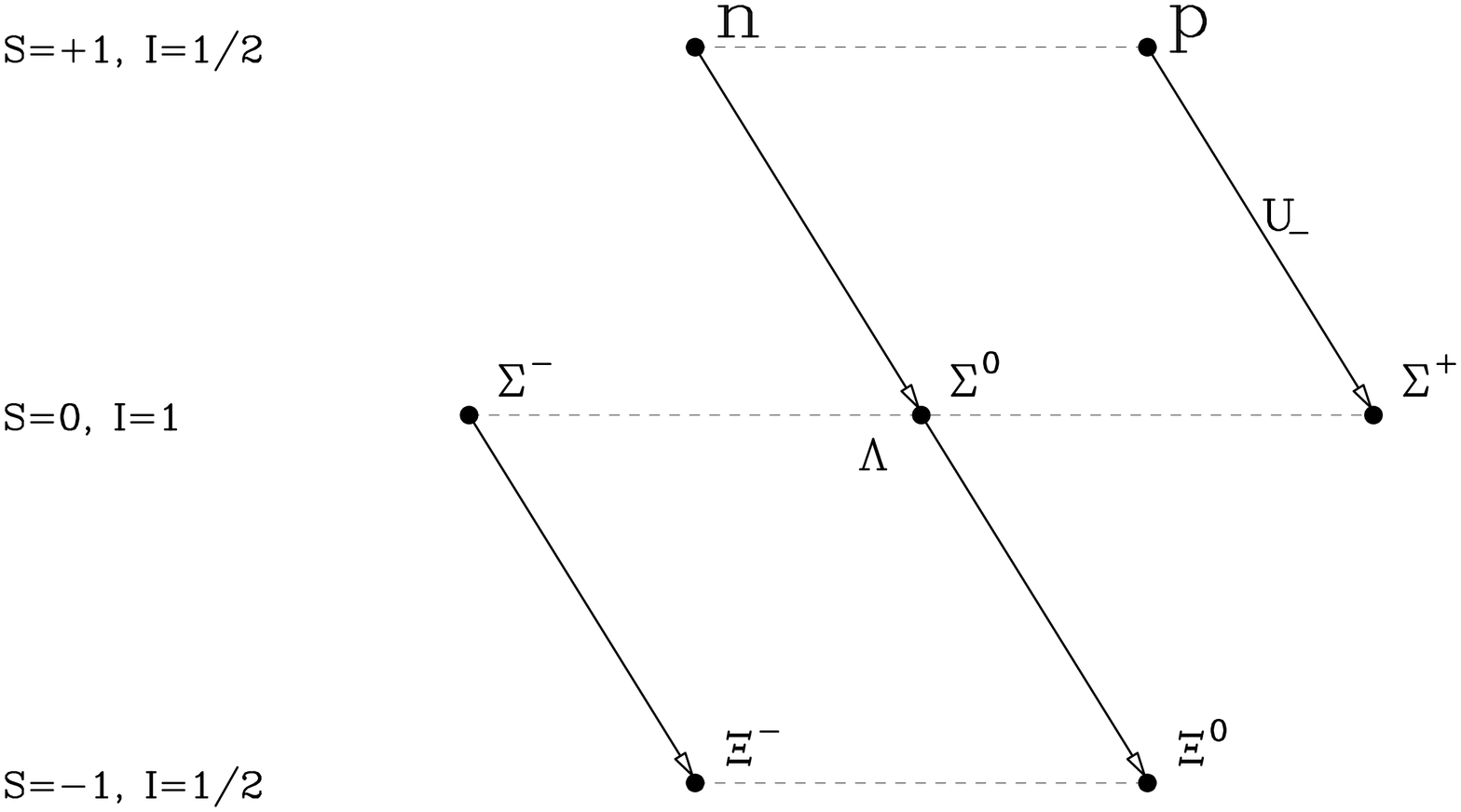}
\qquad
\includegraphics[width=12.5em,clip=true,angle=90]{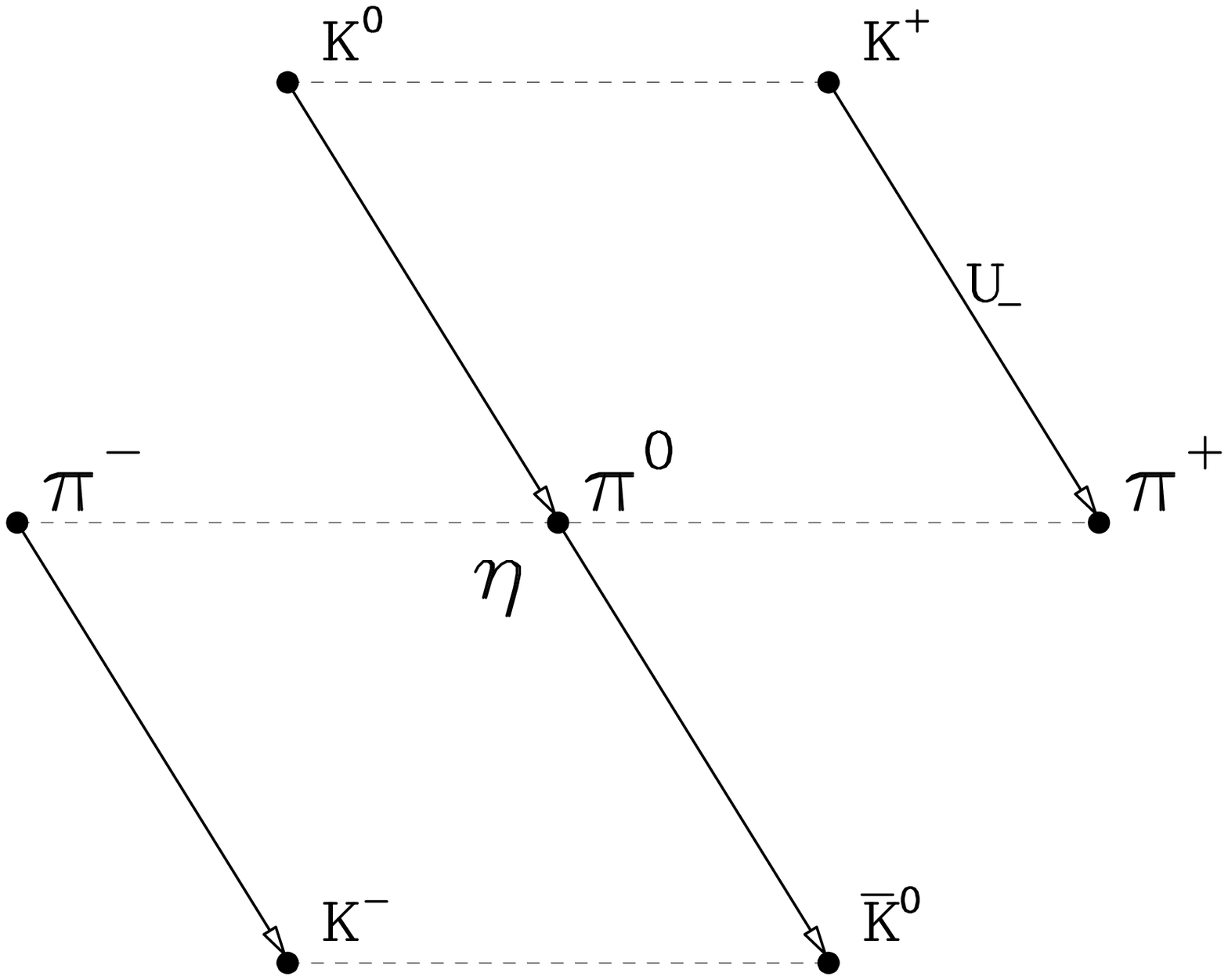}
}
\vskip0.5cm
{\small \em
Fig. 1 (b) Spin-${1\over2}$ baryons and pseudoscalar mesons.
Proton belongs to a $U$-spin doublet with $U={1\over2}$, while neutron
belongs to to a $U$-spin triplet with $U=1$. Similarly,
neutral kaons belong to a $U$-spin triplet
with $U=1$, while charged kaons are members of $U$-spin doublets with
$U={1\over2}$.
\hfill\break
}
} 

 This shows how the photoproduction of
the charged nonstrange member of a $\overline{\bf 10}$ multiplet on a proton is
forbidden by $SU(3)$, while   the photoproduction of the neutral nonstrange member
on a neutron is allowed. However this selection rule holds only
if the
nonstrange resonance is a pure  ($\overline{\bf 10}$) of $SU(3)_f$ with no octet admixture.
But octet-antidecuplet mixing is predicted by $SU(3)_f$ symmetry
breaking. This contradiction can only be resolved my more and better experimental data.

A completely different explanation is needed for
the apparent isospin asymmetry between null \thetap\ photoproduction on protons \cite{g11APS}
vs. clear signal on neutrons (via deuteron target;
e.g.~\cite{Kubarovsky:2003fi},\cite{Nakano2}. It can be resolved if the photon couples
much more strongly to $K^+K^-$ than to $K^o \bar K^o$. A selection rule forbidding
the $\gamma K^o \bar K^o$ vertex follows from conservation of
$G_U$ parity, the $SU(3)_f$
rotation of G-parity with isospin replaced by the $U$-spin.
\cite{Uspin}. The $\gamma$, $K^o$ are $\bar K^o$ are each odd under $G_U$\cite{UspinPhoton},
just as the $\omega$, $\pi^+$ and $\pi^-$ are all odd under ordinary
G parity.
Conservation of $G_U$ parity gives a flavor symmetry
selection rule
forbidding  the $ \gamma K^o \bar K^o$ vertex\cite{gupar}, just as conservation of $G$ parity
forbids $\omega \rightarrow \pi^+ \pi^-$. In the $SU(3)_f$ limit the partial widths of both
transitions are zero
\beq{gammakk}
\Gamma (\gamma \rightarrow \bar K^o K^o) = \Gamma (\omega \rightarrow \pi^+ \pi^-) =0
\end{equation}
This selection rule is confirmed by  the
experimentally observed asymmetry
$\gamma p \rightarrow \Lambda(1520) K^+ \gg \gamma  n \rightarrow \Lambda(1520) K_s$.

A photon which turns into a $G_U$ allowed $K^+K^-$ can make a \thetap\,$K^-$
directly on a neutron, but cannot make a \thetap\ directly on a proton.
A photon which turns into the $G_U$ forbidden $K^o \bar K^o$ can make a
\thetap\,$\bar K^o$ directly on a proton, but cannot make a \thetap\ directly on a neutron.
    How much of the photon appears as $K^+K^-$ and how much as $K^o \bar K^o$
depends upon the amount of $SU(3)_f$ breaking. This is an experimental question
that can be clarified by measuring the neutral
kaons, e.g. comparing $\gamma p \rightarrow p K^+ K^-$ with
the analogous reaction with neutral kaons in the final state.
One can also do an analogous comparison  in $\gamma d$.
The relevant data might in principle be available in the CLAS g11 and
LEPS experiments.

\subsection {\thetap\ photoproduction on proton vs neutron targets}

An isospin analysis of pentaquark production using the selection rule
 forbidding the $\gamma K^o \bar K^o$ vertex also forbids $\gamma p \rightarrow \Theta^+ K_S$ and allows
$\gamma n \rightarrow \Theta^+ K^-$.
Thus $\gamma p \rightarrow \Theta^+ K_S$ can be expected to be suppressed in
comparison  with
$\gamma n \rightarrow \Theta^+ K^-$ if $SU(3)_f$ is not badly
broken.
The validity of the selection rule is tested by the prediction of a
strong $p/n$ asymmetry in
$\gamma N \rightarrow \Lambda(1520) K$.
Recent data on $\Lambda(1520)$ production confirm this picture.
LEPS observes a strong asymmetry in photoproduction of $\Lambda(1520)$ on
proton and neutron \cite{Nakano2}.
They measured both \,
$\gamma p \rightarrow \Lambda(1520) K^+$
and
$\gamma d \rightarrow \Lambda(1520) K N$
and find that the production rate on the deuteron is almost equal to and
even slightly smaller than on the proton. This implies that
\hbox{$\gamma n \rightarrow \Lambda(1520) K_S$} is negligible.

We therefore propose the suppression of the $\gamma K^o \bar K^o$ vertex
as the likely explanation of
non-observation of \thetap\ production on a proton target in the
CLAS g11 experiment \cite{g11APS}. To gain confidence in this explanation
it is important that other experiments confirm the
relative suppression of $\gamma n \rightarrow \Lambda(1520) K$.

In the vector dominance picture the photon is an $SU(3)_f$ octet and a  $U$-spin
scalar combination of the  $\rho$, $\omega$ and $\phi$. In unbroken $SU(3)_f$
they are exactly degenerate and  their contributions to $\gamma \rightarrow K^o
\bar K^o$ cancel exactly.  In the real world $SU(3)_f$ breaking as measured by
vector meson masses is about 25\%-30\%,
so  the relative importance of the $\phi$ component is an open question. Further
implications of this breaking are discussed below.

\mystrut
If the $\gamma K^o \bar K^o$ vertex is indeed strongly suppressed,
then the g11 reaction $\gamma p \rightarrow n K^+ K_S$
proceeds via
$\gamma p \rightarrow K^+K^- p \rightarrow K^+ K_S n$. It
goes via the dominant $\gamma K^- K^+$ coupling and a simple
$K^- p \rightarrow K_S n$
charge exchange.
If this is the dominant mechanism,
there is no possibility of making the $\Theta^+$
in any simple way in the g11 setup.

\mystrut
 The basic physics here is that the photon couples much more strongly
to charged kaons than to neutral kaons and charged kaons can make the
$\Theta^+$ simply on a neutron and not on a proton.

One test of this picture is to compare the photoproduction of isoscalar
baryon resonances with positive and negative strangeness
on proton and neutron targets.

\mystrut
Positive strangeness resonances like the \thetap\ will be produced on neutron
targets and not on protons, while negative strangeness resonances like the
$\Lambda(1520)$ will be produced on proton
targets and not on neutrons.
The favored reactions with charged kaons are:
\beq{allowed}
\gamma p \rightarrow K^+ \Lambda; ~ ~ ~ \gamma n \rightarrow K^- \Theta^+
\end{equation}
The suppressed reactions with neutral kaons are
\beq{suppressed}
\gamma p \rightarrow \bar K^o \Theta^+; ~ ~ ~  \gamma n \rightarrow \bar K^o \Lambda
\end{equation}

\subsection{$SU(3)$ breaking and the $\phi$ component of the photon}

So far we have considered the $\gamma K \bar K$ with unbroken $SU(3)_f$ as the dominant
mechanism for photoproduction of strange baryons.
To investigate $SU(3)_f$  breaking
we introduce the vector dominance picture with broken $SU(3)_f$
and octet-singlet mixing.
The  $\rho$ and $\omega$ components of the photon are still degenerate  but the
$\phi$ is now separate. In this picture for \thetap\ photoproduction the $\bar
s$ strange antiquark is already present in the initial state in the isoscalar
$\phi$ component. The $\rho$ and $\omega$ components contain no strangeness and
can produce the \thetap\  only via the production of an $s \bar s$ pair from
QCD gluons.  How much this extra strangeness production costs is still open.
This cost does not appear in other treatments
\cite{Hosaka}  which involve only pions and ignore the $\phi$ component of
the photon.

\mystrut
An experiment that can check $SU(3)_f$  breaking projects
out the $I=0$ state of a $KN$ state. One example
is the photoproduction on a deuteron \cite{Nakano2}  of the final state
$\Lambda(1520) K^+ n$.

If the strangeness in the reaction comes from the isoscalar $\phi$
component of the photon, the final $KN$ state is required by isospin invariance
to be isoscalar, and the signal is observed against a purely isoscalar
background.
This is not true for the other $K^-pK^+n$ final states  observed
in the same experiment where the  $K^-p$ is not in the $\Lambda(1520)$ and the
effects of a nonresonant $I=1$ background can give very different
results.\footnote{We recall old SLAC experiments
\cite{Boyarski:1970yc,Quinn:1975} which looked at photoproduction of $K^+$-hyperon from
hydrogen and deuterium at 11 and 16 GeV. It would be interesting to
re-examine these data in view of \cite{Nakano2}.}

It is interesting to compare  the reactions
\beq{allowed1}
\gamma p  \rightarrow K^- K^+ p
\end{equation}
\beq{suppressed1}
\gamma p \rightarrow \bar K^o K^o p
\end{equation}
in the neighborhood of the $\phi$ resonance.
At the peak of the $\phi$ the two cross sections should be equal. But on the low mass side of the peak the reaction
(\ref{suppressed1}) should be suppressed relative to the reaction (\ref{allowed1}). Preliminary results support this prediction \cite{amaryan}. An analogous effect has been seen
in the photoproduction of tensor mesons\cite{tensor} in the neighborhood of the hidden strangeness tensor meson 
$f'_2(1525)$

\section {Extension of the diquark-triquark model for
$\Theta^+$ to a nonstrange pentaquark}

\subsection {Review of the dynamics of the $\Theta^+$ diquark-triquark model}

The choice of a pentaquark
wave function is dominated by the short-range hyperfine interaction.

\beq{colspin}
V_{hyp} =-V(\vec \lambda_i \cdot \vec \lambda_j)(\vec \sigma_i \cdot
 \vec \sigma_j)
\end{equation}
where $\vec \lambda$ is N $SU(3)_c$  generator and $\vec \sigma$
is a Pauli spin operator. The sign and magnitude
are normalized by $\Delta$-$N$ mass splitting. The sign shows
that the
$qq$ interaction is attractive
in states symmetric in color and spin and repulsive in antisymmetric states.

The Pauli principle forces two identical fermions
to be in the repulsive antisymmetric color-spin state at short distances
where the wave function must be symmetric in space.
The hyperfine interaction is therefore repulsive between
$uu$ and $dd$  pairs in a nucleon or pentaquark.

The optimum wave function with minimum color-magnetic energy
keeps like-flavor $uu$ and $dd$ pairs apart,  while minimizing the
distance and optimizing the color couplings within the  other pairs.


Thus the diquark-triquark model for the pentaquark has  unusual color structure;
a $\overline{\hbox{\bf 3}}_c$
$ud$ diquark is coupled to a $\hbox{\bf 3}_c$ $ud\bar s$
triquark in a relative $P$-wave to make a state with
$J^P=1/2^+$, $I=0$. The color-magnetic short-range hyperfine
interaction $V_{hyp}$ is dominant for possible binding.
The two pairs of identical quarks $(uu)$ and $(dd$) which have
repulsive short range interactions are separated by placing one
in a diquark and one in the triquark, thus eliminating the short
range repulsion.

\subsection {The nonstrange pentaquark with dominant symmetry-breaking; $m_s - m_d$}

A general pentaquark has two allowed $SU(3)_f$ couplings, an
antidecuplet $\bar {\hbox{\bf 10}}$
and an octet ${\hbox{\bf 8}}$.
Both are found in a system of a diquark in
$\bar {\hbox{\bf 3}}$
and a triquark in $\bar{\hbox{\bf 6}}$ of
$SU(3)_f$.

$6 \times 3 = 18 = 10 + 8 $

There is no  positive strangeness state in the
octet.
Thus only the $\bar {\hbox{\bf 10}}$
antidecuplet is allowed for the positive strangeness $\Theta^+$
with the quark configuration $uudd\bar s$.

Two quark configurations are allowed for a nonstrange pentaquark,
$uuds\bar s$ and $uudd\bar d$. These two configurations are orthogonal
linear combinations of the octet and antidecuplet.
Nonstrange pentaquarks mix octet and antidecuplet like
singlet-octet $\omega-\phi$ mixing.
If the dominant symmetry-breaking interaction is the quark mass difference,  $m_s - m_d$,
the mass eigenstates are the $uuds\bar s$ with hidden strangeness like the $\phi$
and the nonstrange $uudd\bar d$ like the $\omega$.
The flavor-antisymmetry principle
prefers the $uuds\bar s$
configuration which has only one pair of identical quarks. These
can be kept apart in the diquark-triquark model by putting one
u-quark in the diquark and the other in the triquark.

To extend the  diquark-triquark model for a strange
pentaquark to a nonstrange pentaquark we first choose the $uuds\bar s$
with only one pair of identical quarks. We replace the  $ud$ diquark in
$\Theta^+$ by a $us$ diquark and keep the same  $ud\bar s$ triquark.
The system is divided into two color non-singlet clusters in a relative
$P$-wave which
separates pairs of identical flavor
by a distance larger than the range of
the color-magnetic force.

The clusters are kept together by color electric force;
the hyperfine interaction acts only within each cluster.
The $us$ diquark is in the
$\bar {\hbox{\bf 3}}$
of $SU(3)_c$  and $\bar {\hbox{\bf 3}}$ of $SU(3)_f$ to make
a state with $I=1/2, S=0$.

The standard treatment using the $SU(6)$ color-spin algebra shows
the hyperfine interaction stronger
for diquark-triquark
system than $\pi N$ system.
In $SU(3)_f$ symmetry limit
\beq{su3sym}
[V(triquark) + V(diquark)] - [V(K )+ V(N)]=
-{1\over6}(M_\Delta-M_N)  \approx {-}50 {\rm MeV}
\end{equation}

The physics here is simple.
The spin-zero diquark is the same as the diquark
in a $\Lambda$ with the same hyperfine energy as
nucleon.
The triquark with one  quark coupled with a $\bar s$ antiquark to spin
zero has same the hyperfine energy as a kaon but no interaction with the
other quark.
The triquark coupling allows $\bar s$ antiquark to interact with
both $u$ and $d$ quarks and gain hyperfine energy with respect to the case
of the kaon.

An isolated triquark is
not color singlet. The triquark
color charge is neutralized by the diquark.

With first order symmetry breaking  $M(\Lambda) \not= M(N)$ and
the mass of the $\pi N$
pentaquark is predicted to be higher than
$\Theta^+$ mass by  $M(\Lambda) - M(N)$

\beq{pinmass}
M(\pi N)_{pred} =M(\Theta^+) + M(\Lambda) - M(N) = 1540 + 180 = 1720;
 ~ ~ ~ M(\pi N)_{exp} = 1680
\end{equation}

\centerline {\bf
Not bad for such a crude calculation}
\vskip1cm

The $uudd\bar s$ pentaquark is a really complicated five-body
system. Flavor antisymmetry suggests that the commonly used
bag or single-cluster models may be correct to treat normal
hadrons but are not adequate for multiquark systems.
These models have identical pair correlations for all pairs in the
system.
They miss the flavor antisymmetry which requires different short-range
pair correlations between pairs with the same flavor and pairs with different flavors.

Had it not been for the cost of the $P$-wave excitation,
the triquark-diquark system would be somewhat more
bound than a pion and a nucleon.
The diquark and triquark will have a color electric interaction
between them which is identical to the quark-antiquark interaction in a meson.

If we neglect the $P$-wave
excitation energy and the finite sizes of the diquark and triquark we can compare
this system with analogous mesons. We can use the effective quark masses
 that
fit the low-lying mass spectrum
to find a very rough estimate
$$
m_{diq} =  720\ \hbox{MeV},\qquad
m_{triq}=  1260\ \hbox{MeV},\qquad
m_r(di\hbox{-}tri)= 458\ \hbox{MeV}
\,.
$$
where $m_{diq}$ and
$m_{triq}$ denote the
effective masses of the diquark and triquark, and $m_r(di\hbox{-}tri)$ denotes
the reduced
mass for the relative motion of the diquark-triquark system.

A crucial observation has been that the
diquark-triquark system may not exist in
a relative S-wave.
This is because in $S$-wave the hyperfine interaction acts not
only within the clusters but also between them. The repulsive terms may then win and
the would be $S$-wave gets rearranged into the usual meson-baryon system. The situation
is different in a $P$-wave, because then the diquark and the triquark are separated
by an angular momentum barrier and the color-magnetic interactions operate only
within the two clusters.

\section*{Summary and Conclusions}

We discuss the implications of the suggested interpretation  of the recently
reported narrow $\pi N$ resonance (width $\approx$ 25 MeV at 1680 MeV)
as a pentaquark in the same multiplet as the $\Theta^+$.
We consider a diquark-triquark pentaquark model involving a recoupling of
the five quarks into a diquark-triquark system in non-standard color
representations. We estimate the mass using
the simple generalized Sakharov-Zeldovich mass
formula which holds with a single set of effective quark mass values for
all ground state mesons and baryons having no more than one  strange or
heavy quark.

Our rough numerical estimate puts the $\pi N$ resonance in a nonstrange
pentaquark with a mass 180 MeV higher than the $\Theta^+$ mass by $M(\Lambda) - M(N)$.
This gives a pentaquark mass of 1720 MeV, sufficiently close to 1680 MeV to indicate that this
approach deserves further more accurate investigation.

\section*{Acknowledgements}
We thank Moskov Amaryan, Atsushi Hosaka, Uri Karshon, Takashi Nakano and Maxim Polyakov
for discussions. We are grateful to the organizers of the NNR workshop for
creating a stimulating scientific environment.

%
\catcode`\@=11 
\def\references{
\ifpreprintsty \vskip 10ex
%
\hbox to\hsize{\hss \large \refname \hss }\else
\vskip 24pt \hrule width\hsize \relax \vskip 1.6cm \fi \list
{\@biblabel {\arabic {enumiv}}}
{\labelwidth \WidestRefLabelThusFar \labelsep 4pt \leftmargin \labelwidth
\advance \leftmargin \labelsep \ifdim \baselinestretch pt>1 pt
\parsep 4pt\relax \else \parsep 0pt\relax \fi \itemsep \parsep \usecounter
{enumiv}\let \p@enumiv \@empty \def \theenumiv {\arabic {enumiv}}}
\let \newblock \relax \sloppy
 \clubpenalty 4000\widowpenalty 4000 \sfcode `\.=1000\relax \ifpreprintsty
\else \small \fi}
\catcode`\@=12 


\begin{thebibliography}{99}

\bibitem{Arndt:2003ga}
  R.~A.~Arndt, Y.~I.~Azimov, M.~V.~Polyakov, I.~I.~Strakovsky and
R.~L.~Workman,
  Phys.\ Rev.\  C {\bf 69}, 035208 (2004)
  [arXiv:nucl-th/0312126].

\bibitem{Nakano:2003qx}
T.~Nakano {\it et al.}  [LEPS Coll.],
Phys.\ Rev.\ Lett.\  {\bf 91}, 012002 (2003), hep-ex/0301020.

\bibitem{hicks}
For an updated general review see Ken Hicks, hep-ex/0412048
and hep-ex/0501018.
\bibitem{jenmalt}
For a review of the considerable theoretical literature on pentaquark models
and an in-depth discussion, see
Byron K. Jennings and Kim Maltman,
Phys. Rev. {\bf D69} (2004) 094020,
hep-ph/0308286.

\bibitem{NewPenta}
M. Karliner and H.J. Lipkin,
Phys.\ Lett.\ B {\bf 575} (2003) 249,
hep-ph/0307243.

\bibitem{cryptopen}
M. Karliner and H.J. Lipkin,
Phys.\ Lett.\ B {\bf 597} (2004) 309,
hep-ph/0405002.

\bibitem{ichepproc}
H.J. Lipkin, hep-ph/0501209 and
Proc. 32nd Int. Conf. on High Energy Physics,
Beijing, China, 16-22 August 2004,
Hesheng Chen et al., Eds.,
World Scientific, 2005.

\bibitem{Hosaka} Seung-Il Nam,
Atsushi Hosaka and Hyun-Chul Kim,
Genshikaku Kenkyu {\bf 49} (2005) 53,
hep-ph/0502143 and hep-ph/0503149.

\bibitem{g11APS}
R. De Vita, reporting preliminary results from CLAS g11 experiment,
talk at at APS meeting, Tampla, Florida, April 16-19, 2005,
\hfill\break
{\footnotesize\tt
www.jlab.org/div\_dept/physics\_division/talks/Background/Hall\_B/DeVita\_aps2005.ppt
}\ .

\bibitem{Kubarovsky:2003fi}
  V.~Kubarovsky {\it et al.}  [CLAS Collaboration],
  Phys.\ Rev.\ Lett.\  {\bf 92} (2004) 032001
  [Erratum-ibid.\  {\bf 92} (2004) 049902]
  [arXiv:hep-ex/0311046].

\bibitem{Nakano2}T. Nakano,
private communication. More details to be presented at
\hfill\break
{\tt  http://www.phy.pku.edu.cn/{\URLtilde}qcd/}\ .

\bibitem{uspin}
M. Karliner and H.J. Lipkin,
arXiv:hep-ph/0506084

\bibitem{Nam} Seung-Il Nam,
Atsushi Hosaka and Hyun-Chul Kim,
hep-ph/0505005

\bibitem{gupar} Harry J. Lipkin, Phys. Rev. Lett. {\bf 31} (1973) 656

\bibitem{Uspin}S. Meshkov, C.A Levinson, and
H.J. Lipkin, Phys. Rev. Lett. {\bf 10} (1963) 361.
For more recent applications of U spin see
Michael Gronau and Jonathan L. Rosner,
Phys.Lett. B500 (2001) 247 hep-ph/0010237
and references therein.

\bibitem{UspinPhoton}
C.A Levinson, H,J. Lipkin and S. Meshkov
Phys. Lett. {\bf 7}, 8l (1963).


\bibitem{Boyarski:1970yc}
A.~Boyarski et al.,
Phys.\ Lett.\ B {\bf 34} (1971) 547.

\bibitem{Quinn:1975}D.J. Quinn et al
 Phys. \ Rev. \ Lett. {\bf 34} (1975) 543.
\bibitem{amaryan} Moskov Amaryan, private communication
\bibitem{tensor} ZEUS Collaboration, S.Chekanov et al Phys. \ Rev. \ Lett. {\bf 101} (2008) 12003 


\bibitem{HicksPC}
K. Hicks, private communication.

\end{thebibliography}
\end{document}